\documentclass[10pt,journal]{IEEEtran}

\ifCLASSINFOpdf
 
\else
 
\fi

\usepackage[acronym,hyperref]{mycommon}
\usepackage{url}
\usepackage{tikz}
\usetikzlibrary{calc}

\newcommand{\positiontextbox}[4][]{%
	\begin{tikzpicture}[remember picture,overlay]
		\node[inner sep=3pt, fill=yellow,align=left,draw,line width=1pt,#1] at ($(current page.north west) + (#2,-#3)$) {\parbox{.80\paperwidth}{#4}};
	\end{tikzpicture}%
}

\begin{document}
%
\title{Analog Beamforming using Time-Modulated Arrays with Digitally Preprocessed Rectangular Sequences}
%
%
%

\author{Roberto Maneiro-Catoira,~\IEEEmembership{Member,~IEEE,}
        Julio Br\'egains,~\IEEEmembership{Senior~Member,~IEEE,}\\
        Jos\'e A. Garc\'ia-Naya,~\IEEEmembership{Member,~IEEE,}
        and~Luis Castedo,~\IEEEmembership{Senior~Member,~IEEE}
\thanks{%
This work has been funded by the Xunta de Galicia (ED431C 2016-045, ED341D R2016/012, ED431G/01), the Agencia Estatal de Investigaci\'on of Spain (TEC2013-47141-C4-1-R, TEC2015-69648-REDC, TEC2016-75067-C4-1-R) and ERDF funds of the EU (AEI/FEDER, UE).
	
The authors are with the University of A Coru\~na, Spain. E-mail: \{roberto.maneiro, julio.bregains, jagarcia, luis\}@udc.es}
\thanks{}}

%
%

\markboth{IEEE ANTENNAS AND WIRELESS PROPAGATION LETTERS}%
{}
%



\maketitle

\acrodef{ADC}[ADC]{Analog to Digital Converter}
\acrodef{AWGN}[AWGN]{Additive White Gaussian Noise}
\acrodef{ASK}[ASK]{Amplitude-Shift Keying}
\acrodef{BF}[BF]{Beamforrming}
\acrodef{BER}[BER]{Bit Error Ratio}
\acrodef{DAC}[DAC]{Digital to Analog Converter}
\acrodef{DC}[DC]{Direct Current}
\acrodef{DoA}[DoA]{Direction of Arrival}
\acrodef{DSB}[DSB]{Double Side Band}
\acrodef{FSK}[FSK]{Frequency-Shift Keying}
\acrodef{FT}[FT]{Fourier Transform}
\acrodef{GA}[GA]{Genetic Algorithm}
\acrodef{ISI}[ISI]{Inter-Symbol Interference}
\acrodef{LBFN}[LBFN]{Linear Beamforming Network}
\acrodef{NPD}[NPD]{Normalized Power Density}
\acrodef{PSK}[PSK]{Phase-Shift Keying}
\acrodef{PSO}[PSO]{Particle Swarm Optimization}
\acrodef{QAM}[QAM]{Quadrature Amplitude Modulation}
\acrodef{RF}[RF]{Radio Frequency}
\acrodef{SA}[SA]{Simulated Annealing}
\acrodef{SLL}[SLL]{Side-Lobe Level}
\acrodef{SR}[SR]{Sideband Radiation}
\acrodef{SNR}[SNR]{Signal-to-Noise Ratio}
\acrodef{SSB}[SSB]{Single-Sideband}
\acrodef{SWC}[SWC]{Sum of Weighted Cosines}
\acrodef{TM}[TM]{Time Modulation}
\acrodef{TMA}[TMA]{Time-Modulated Array}
\acrodef{VFDF}[VFDF]{Variable Fractional Delay Filter}
\acrodef{VGA}[VGA]{Variable Gain Amplifier}
\begin{abstract}
Conventional time-modulated arrays are based on the application of variable-width periodical rectangular pulses (easily implemented with radio frequency switches) to the individual antenna excitations. However, a serious bottleneck arises when the number of exploited harmonic beams increases. In this context, the modest windowing features of the rectangular pulses produce an inflexible and ineffective harmonic beamforming. The use of other pulses like sum of weighted cosines partially solves these issues at the expense of introducing additional non-timing variables. We propose the discrete-time preprocessing of rectangular pulses before being applied to the antenna to accomplish an agile, efficient and accurate harmonic beamforming, while keeping the simplicity of the hardware structure.  
\end{abstract}


\acresetall

%
\IEEEpeerreviewmaketitle

\positiontextbox{11cm}{27cm}{\footnotesize \textcopyright 2018 IEEE. This version of the article has been accepted for publication, after peer review. Personal use of this material is permitted. Permission from IEEE must be obtained for all other uses, in any current or future media, including reprinting/republishing this material for advertising or promotional purposes, creating new collective works, for resale or redistribution to servers or lists, or reuse of any copyrighted component of this work in other works. Published version:
	\url{https://doi.org/10.1109/LAWP.2018.2797971}}

\section{Introduction}
\IEEEPARstart{T} {he} design of \ac{BF} architectures for advanced multi-antenna systems is unavoidably constrained by the trade-off between flexibility, performance and simplicity. In this respect, full-blown digital \ac{BF} --which requires one \ac{RF} chain per antenna-- uses discrete baseband processing to synthesize the \ac{BF} weights, offering unbeatable flexibility and performance at the expense of increased complexity, cost and power consumption. Accordingly, digital \ac{BF} is attractive when the performance outweighs mobility but it becomes unpractical for large-scale antenna systems where the physical space is a handicap.
 
\begin{figure}[!t]
	\centering
	\includegraphics[width=0.7\columnwidth]{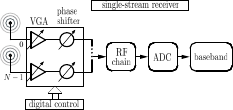}
	\caption{Receive analog beamformer consisting of a linear array with $N$ isotropic elements, implemented using a network of digitally controlled phase shifters and \acp{VGA}.This hardware scheme is not suitable for multi-stream nor multi-user communications.}
	\label{fig:Typical_ABF}
\end{figure}

Alternatively, analog \ac{BF} is a simple and cost-effective method for generating high-gain pencil beam patterns (especially helpful to mitigate the enormous path losses at high frequencies) but is subject to a series of restrictions in terms of flexibility. Indeed, analog \ac{BF} is based on a simple hardware scheme: an antenna array whose individual elements are connected via phase shifters and \acp{VGA} (usually supporting digital control) to a single \ac{RF} chain, as shown in \cref{fig:Typical_ABF} for the case of a receive analog beamformer. Nevertheless, its performance is limited by a number of constraints: (1) the use of quantized phase shifts jeopardizes an accurate steering of the beams and/or the nulls; and, conversely, the higher the resolution, the higher the power consumption; (2)  phase shifters --especially the passive ones at high frequencies-- introduce large insertion losses; and (3) a single beamformer (\cref{fig:Typical_ABF}) only supports a single-stream reception. Therefore, an $L$-stream fully analog \ac{BF} receiver using an array of $N$ elements requires $L$ RF front-ends and a \ac{LBFN} with $L\times N$ phase shifters \cite{Maneiro2017_a}. Usually, $N\gg L$ (especially with large-scale antennas) and the resulting hardware scheme for this analog \ac{BF} is still simpler than the digital one.

With the previous pros and cons in mind, hybrid beamforming architectures benefit from the advantages of both analog (generating sharp beams with a simpler hardware) and digital (providing flexibility and performance) domains. In this context, the analog \ac{BF} will continue to play an increasingly important role. This letter aims to deepen the aforementioned threefold trade-off but focusing on a particular analog \ac{BF} method, namely the \ac{TMA} \ac{BF}. 

\begin{figure*}[t]
	\centering
	\includegraphics[width=1.6\columnwidth]{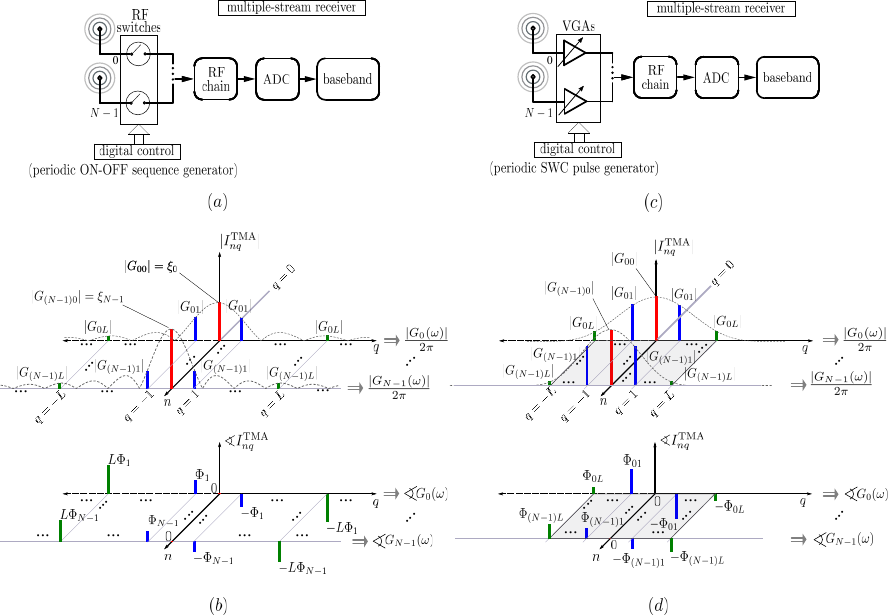}
	\caption{(a) Multi-stream receiver based on \ac{BF} with a conventional \ac{TMA}. (b) Amplitude and phase representation of the dynamic excitations of the \ac{TMA} in \cref{fig:existing_TMAs}(a). The plot reveals: (1) a  modest harmonic windowing behavior when $L>1$, (2) the presence of duplicated-specular radiation diagrams due to the negative harmonics with same amplitude and opposite phase, (3) the scanning inability for the $q=0$ beam due to the zero phases, (4) the time-linear control of the amplitudes at $q=0$ where $\xi_n=\tau_n/T_0$, (5) the proportionality between the phases of harmonics with different order. (c) Multi-stream receiver based on a \ac{TMA} \ac{BF} with \ac{SWC} pulses \cite{Maneiro2017_b}. (d) Amplitude and phase representation of the dynamic excitations of the \ac{TMA} in \cref{fig:existing_TMAs}(c).}
	\label{fig:existing_TMAs}
\end{figure*}

Two different \ac{TMA} implementation structures have been proposed as alternatives to the one in \cref{fig:Typical_ABF}: 1) conventional \acp{TMA} with \ac{RF} switches \cite{Rocca2014,Maneiro2017_a} (see \cref{fig:existing_TMAs}(a)), in which rectangular pulses become a serious bottleneck in terms of flexibility and efficiency, specially when the number of exploited harmonic beams increases; and 2) the so-called Enhanced \acp{TMA} proposed in \cite{Maneiro2017_b} (see \cref{fig:existing_TMAs}(c)), based on \acp{VGA} governed by \ac{SWC} pulses, which solve completely the flexibility problem, but only partially the efficiency issues caused by specular beams corresponding to the negative harmonics and a fundamental mode beam with scanning inability. In this letter we propose a novel \ac{TMA} structure that overcomes those flexibility and efficiency issues.

\section{TMA Beamforming: Strengths and Weaknesses }\label{sec:TMA beamforming structures}

By replacing the phase shifters and the \acp{VGA} in the scheme in \cref{fig:Typical_ABF} with digitally controlled \ac{RF} switches --and hence considering a unitary and uniform static distribution of the array excitations-- we obtain the conventional \ac{TMA} receiver shown in \cref{fig:existing_TMAs}(a). The application of periodical ($T_0$) rectangular pulses $g_n(t)$ --easily adjustable in width ($\tau_n$) and delay ($\delta_n$) \cite{Rocca2014}-- to the individual antenna excitations causes the appearance of radiation patterns at the harmonic frequencies $\omega_q=\omega_c + q\omega_0$, with $q \in \mathbb{Z}\setminus\{0\}$, $\omega_0=2\pi/T_0$, and $\omega_c$ being the carrier frequency. Accordingly, the \ac{BF} can be performed over the harmonic frequencies $\omega_q$ with $q \in \{ \pm1,\pm2,\cdots,\pm L\}$, where $L$ is the order of the highest exploited harmonic. The array factor --with the term $\expe{j\omega_qt}$ explicitly included-- corresponding to the harmonic frequency $\omega_q$ is given by $F_q^{\text{TMA}}(\theta,t)=\expe{j\omega_qt}\sum_{n=0}^{N-1} I_{nq}^{\text{TMA}} \expe{jkz_n\cos\theta}$ \cite{Maneiro2017_a}, where $z_n$ represents the $n$-th array element position on the $z$ axis,  $\theta$ is the angle with respect to the main axis of the array, $k$ is the wavenumber, and $I_{nq}^{\text{TMA}}$ is the complex-valued representation of the per-antenna time-modulated current excitation. 

Provided that $\omega_0>B$ ($B$ is the signal bandwidth), and that the working harmonic patterns are designed (by suitably selecting $\tau_n$ and $\delta_n$) fulfilling spatial orthogonality, the first $L$ positive harmonics can be exploited to receive $L$ different incoming signals at $\omega_c$  \cite{Maneiro2017_a}.  Such signals are received over  different \ac{TMA} harmonic beams and inherently translated into the corresponding $\omega_q$, occupying a total bandwidth of $L\omega_0$. This multi-stream receiver can be implemented with the extraordinarily simple hardware scheme in \cref{fig:existing_TMAs}(a), with the remarkable benefit of needing a single \ac{RF} front-end (with a higher bandwidth and a faster \ac{ADC} than that in \cref{fig:Typical_ABF}). However, such a multi-stream capacity is severely limited. In order to make more visible these constraints, let us firstly consider the following  property of the dynamic excitations  which is crucial for the  harmonic \ac{BF} design: $I_{nq}^{\text{TMA}}=G_{nq}=G_n(q\omega_0)/2\pi$, with $G_{nq}$ being the Fourier series coefficients of  $g_n(t)$, and $G_n(\omega)$ its Fourier transform. $G_n(\omega)$ is a discrete spectrum with impulses at integer multiples of $\omega_0$  and whose envelope is the Fourier transform of the aperiodic basic pulse that conforms a single period of $g_n(t)$. 

Accordingly, we can represent both the amplitude and the phase of $I_{nq}^{\text{TMA}}$ (functions of $\tau_n$ and/or $\delta_n$) in a two-dimensional grid with the variables $n$ and $q$. Such a portrayal of the TMA for the case of the receiver in \cref{fig:existing_TMAs}(a) is illustrated in \cref{fig:existing_TMAs}(b) and reveals a series of critical aspects that directly impact on the \ac{BF} performance. Thus, with respect to the \ac{TMA} power efficiency, and due to the poor windowing features of the $\sinc$ envelope\footnote{More specifically, it shows a modest maximum side lobe level of $-13$\,dB and a first order asymptotic decay.} of $|I_{nq}^{\text{TMA}}|$, the spectral energy is not efficiently distributed among the working harmonics, causing that, for values of $L>1$, the antenna efficiency be significantly deteriorated \cite{Maneiro2017_a}. Moreover, as $g_n(t)$ is a real-valued signal, its Fourier coefficients verify $G_{nq}=G^*_{n(-q)}$, leading to  $|F_q^{\text{TMA}}(\theta)|^2=|F_{-q}^{\text{TMA}}(180^{\circ}-\theta)|^2$, thus synthesizing a pair of diagrams for $q$ and $-q$, which are symmetric with respect to $\theta=90^{\circ}$. In general, such specular ``$-q$'' diagrams are not useful, spoiling the \ac{TMA} efficiency even more. This weakness motivated the work in \cite{Yao_2015}, in which a \ac{SSB} \ac{TMA} is modeled with exclusive control over the phase (and hence without \ac{SLL} control) and capable of handling a single beam ($L=1$).

The second critical aspect in the \ac{BF} design is the \ac{TMA} beam-steering efficiency. Remarkably, the efficiency term is not related to the angular accuracy (precisely an advantage of the \acp{TMA} which allows for a continuous-time phase control), but to the ability to point each harmonic beam towards the corresponding desired direction. In this regard, we could also include the aforementioned specular ``$-q$'' beams, which point towards spatial directions probably not usable, but there are other relevant hindrances (see \cref{fig:existing_TMAs}(b)): 1) the scanning inability of the \ac{TMA} fundamental pattern ($q=0$) due to the unavoidable zero phases of its excitations ($\sphericalangle I_{n0}^{\text{TMA}}=0$), and 2) the proportionality between the phases of harmonics of different order, quantitatively, $\Phi_{nq}=-q\Phi_{n1}$, which implies a lack of flexibility to independently steer harmonic beams with different order when $L>1$. 

The \ac{TMA} beamformer in \cref{fig:existing_TMAs}(c), employs digitally controlled \acp{VGA} governed by \ac{SWC} pulses instead of \ac{RF} switches \cite{Maneiro2017_b}. Such a scheme solves the efficiency problem with regard to an appropriate windowing of the working harmonics (see \cref{fig:existing_TMAs}(d)) at the expense of introducing non-timing parameters, but still suffers from the presence of the negative harmonic patterns. Remarkably, the system accomplishes full independence between harmonic beams with different orders thanks to the synthesis of independent dynamic excitation phases $\sphericalangle I_{nq}^{\text{TMA}}$ (see \cref{fig:existing_TMAs}(d)) through the periodic convolution with an auxiliary periodic signal \cite{Maneiro2017_b}.

\begin{figure}[t]
	\centering
	\includegraphics[width=0.7\columnwidth]{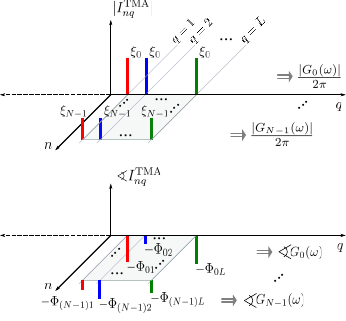}
	\caption{Amplitude and phase representation of the dynamic excitations of the proposed \ac{SSB} \ac{TMA} beamformer.}
	\label{fig:design_goal}
\end{figure}

\section{TMA Beamforming Design}\label{sec:TMA beamforming design}

In order to solve the \ac{BF} pending issues detailed above, we propose to address the \ac{TMA} design taking a desired distribution of the dynamic excitations $I_{nq}^{\text{TMA}}$ as a departure point. In this respect, and as a first approach, our aim is to synthesize $N$ identical-in-shape pencil beams (reconfigurable in terms of \ac{SLL}) with fully-independent steering capacity. Thus, we consider the $I_{nq}^{\text{TMA}}$ distribution in \cref{fig:design_goal}. Such a distribution constitutes a bidimensional frequency comb located exclusively in the positive\footnote{Note that the dynamic excitations lines at $q=0$ are also removed.} harmonics with the aim of synthesizing a multi-stream receiver based on a \ac{SSB} \ac{TMA}. We also highlight the time-linear control of the amplitudes, i.e., $|I_{nq}^{\text{TMA}}|=\tau_n/T_0=\xi_n$, implying a simpler reconfiguration and a more accurate control of the \ac{BF}. Regarding $\sphericalangle I_{nq}^{\text{TMA}}$, notice that they are independent as in \cref{fig:existing_TMAs}(d). 
 \begin{figure}[t]
 	\centering
 	\includegraphics[width=0.7\columnwidth]{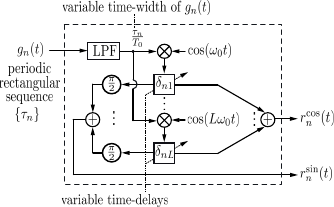}
 	\caption{Block diagram of the preprocessing of the periodic rectangular pulse to synthesize the two quadrature periodic pulses which govern the $n$-th antenna element of the proposed \ac{SSB} \ac{TMA} in \cref{fig:SSB TMA HW}.}
 	\label{fig:Preprocessing of the DC}
 \end{figure}
  \begin{figure}[t]
 	\centering
 	\includegraphics[width=0.7\columnwidth]{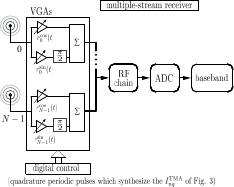}
 	\caption{Proposed multi-stream receiver based on \ac{BF} with an \ac{SSB} \ac{TMA} with preprocessed rectangular sequences.}
 	\label{fig:SSB TMA HW}
 \end{figure}
 
 Once fixed the desired dynamic excitations, we must design the periodic pulses (governed exclusively by time parameters) as well as the \ac{TMA} structure capable of synthesizing such time-controlled excitations. The idea is as simple as effective: to filter out the \ac{DC} component of each periodic rectangular pulse $g_n(t)$ --corresponding to $G_{n0}=I_{n0}^{\text{TMA}}=\xi_n$-- which is the one with a time-linear controlled amplitude, and immediately afterwards, apply a multicarrier complex modulation to such a \ac{DC} signal (see \cref{fig:Preprocessing of the DC}) in order to shift both time-linear control of the amplitudes\footnote{Other research lines, out of the scope of \ac{TMA} philosophy, propose that the amplitude tapering of the pulses be non-timing parameters \cite{He2013URSI}.} and time control of the phases to the spectral lines at the frequencies $q\omega_0$. In \cref{fig:Preprocessing of the DC}, $\delta_{nq}$ are time delays\footnote{Time delays can be accurately implemented in the digital domain through \acp{VFDF} \cite{Soo_2013}.} that introduce a phase shift in the corresponding dynamic excitation given by $\sphericalangle I_{nq}^{\text{TMA}}=-\Phi_{nq}=-q\omega_0\delta_{nq}$. The quadrature periodic pulses obtained by this procedure allow for implementing the \ac{SSB} features of the \ac{TMA}, and are respectively given by (see \cref{fig:Preprocessing of the DC}): $r_n^ {\cos}(t)=\xi_n\sum_{q=1}^{L}\cos(q\omega_o t-\Phi_{nq})$ and  $r_n^ {\sin}(t)=\xi_n\sum_{q=1}^{L}\sin(q\omega_o t-\Phi_{nq})$.
 
 The proposed multi-stream receiver structure based on an \ac{SSB} \ac{TMA} is shown in \cref{fig:SSB TMA HW}. Let us consider that a signal $s(t) \in \mathbb{R}$ at $\omega_c$ impinges on the \ac{TMA}. For the sake of simplicity, we will work with its complex-valued representation $\tilde{s}(t)=u(t)\expe{j\omega_ct} \in \mathbb{C}$, being $u(t) \in \mathbb{C}$ the equivalent baseband signal of $s(t)$, i.e., $s(t)=\Re\{\tilde{s(t)}\}$. According to \cite{Maneiro_2014} and \cref{fig:SSB TMA HW}, the corresponding signal at the output of the \ac{TMA} is
\begin{equation}
 \tilde{s}^{\text{TMA}}(t,\theta)=u(t)\expe{j\omega_ct}\sum_{n=0}^{N-1}(r_n^{\cos}(t)+jr_n^ {\sin}(t))\expe{jkz_n\cos\theta}, 
\end{equation} 
whose Fourier transform --after immediate calculations with the simple Fourier transforms of $r_n^ {\cos}(t)$ and $r_n^{\sin}(t)$, where the $\delta(\omega+q\omega_o)$ terms are canceled out-- is
\begin{equation}
\tilde{S}^{\text{TMA}}(\omega,\theta)=\tfrac{U(\omega-\omega_c)}{2\pi}\ast\sum_{n=0}^{N-1}\sum_{q=1}^{L}\xi_n\delta(\omega-q\omega_o)\expe{-j\Phi_{nq}}\expe{jkz_n\cos\theta}, 
\end{equation} 
with "$\ast$" being the convolution operator. Turning back to the time domain through the inverse Fourier transform we have
\begin{align}
\tilde{s}^{\text{TMA}}(t,\theta)&=2u(t)\sum_{q=1}^{L}\sum_{n=0}^{N-1}\xi_n\expe{-j\Phi_{nq}}\expe{jkz_n\cos\theta}\expe{j\omega_qt}\notag\\
&=2u(t)\sum_{q=1}^{L}\expe{j\omega_qt}\sum_{n=0}^{N-1}I_{nq}^{\text{TMA}}\expe{jkz_n\cos\theta}\notag\\
&=2u(t)\sum_{q=1}^{L}F_q^{\text{TMA}}(\theta,t)=2u(t)F^{\text{TMA}}(\theta,t),
\end{align} 
where it is demonstrated that $I_{nq}^{\text{TMA}}=\xi_n\expe{-j\Phi_{nq}}$ which corresponds exactly to the target distribution of \cref{fig:design_goal}.  

\section{Simulated Example: A Brief Comparison}\label{sec:Simulation}
\begin{figure}[t]
	\centering
	\includegraphics[width=0.8\columnwidth]{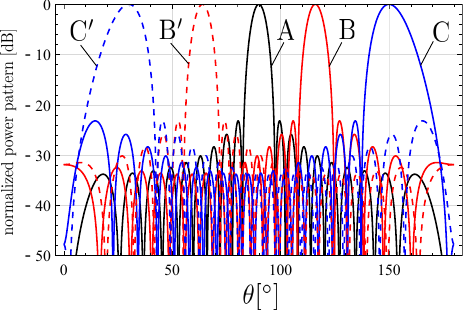}
	\caption{Beams A, B, C, B', and C' are synthesized both with a \ac{TMA} with switches and a \ac{TMA} with \ac{SWC} pulses over the harmonics $q=\{0, 1, 2, -1, -2\}$. The proposed \ac{SSB} \ac{TMA} only generates the useful beams A, B, C over the harmonics $q=\{1, 2, 3\}$.}
	\label{fig:example}
\end{figure}

In this example we take as a reference the radiation pattern in \cite[Fig.~11]{Rocca2014}, synthesized with a \ac{TMA} with switches of $N=20$ elements and $z_n=\lambda/2$, which performs \ac{BF} exploiting the beams $q=\{0, 1, 2\}$ corresponding to patterns A, B, and C in \cref{fig:example}. Notice that specular beams at $q = \{-1, -2\}$ (not shown in \cite{Rocca2014} but introduced in \cref{fig:example} as patterns B' and C') are spontaneously generated. The corresponding pulse sequences (see \cite[Fig.~11(c)]{Rocca2014}) applied to the static uniform excitations, are obtained through a \ac{PSO} algorithm. In this \ac{TMA} structure, the beam $q=0$ has no steering capacity and the theoretical antenna efficiency (note that the hardware efficiency \cite{Maneiro2017_a} is not considered because it depends on the specific selected devices) is calculated as the quotient between the useful and the total mean received powers: $\eta(L)=P_U^{\text{TMA}}/P_R^{\text{TMA}}=\sum_{q=0}^{L-1}p_q/\sum_{q=-\infty}^{\infty}p_q$, where $p_q$ is the mean received power over the harmonic $q$ \cite{Maneiro2017_a}. Note that $1-\eta(L)$ corresponds to the power percentage wasted by the unexploited harmonics. In this example $\eta(L=3)=30.78\,\%$. 

The same pattern can be obtained using identical \ac{SWC} pulses whose cosine weights \cite{Maneiro2017_b} are $a_{n2}= a_{n1} = 2a_{n0} = 1/5$ and with $\xi_n$ corresponding to a normalized Gaussian pattern with a standard deviation of $2/3$, leading to $\eta(L=3)=88.30\,\%$. We can synthesize the useful beams A, B, and C (see  \cref{fig:example}) over the harmonics $q=\{1, 2, 3\}$ with the proposed \ac{SSB} \ac{TMA}, using the same $\xi_n$ as for the \ac{SWC} case, yielding $\eta(L=3)=100.00\,\%$ and allowing for steering capacity for all the involved beams. Finally, we quantify the directivity of pattern A for the three approaches, i.e., switches, \ac{SWC} pulses, and the proposed method, arriving at $2.76$\,dBi, $11.58$\,dBi, and $13.82$\,dBi, respectively.

\section{Conclusions}\label{sec:Conclusions}
We have proposed a novel strategy for multiple-beam harmonic \ac{BF} design with \acp{TMA} based on the preprocessing of rectangular pulses with the following features: (1) exclusive use of time parameters (agility), (2) independent scanning of the harmonic beams (flexibility), (3) \ac{SSB} behavior leading to an improved directivity as well as steering ability of all the working beams (efficiency), and (4) time-linear control of the magnitudes of the dynamic excitations without using synthesis optimization algorithms (simplicity).

\ifCLASSOPTIONcaptionsoff
  \newpage
\fi



\bibliographystyle{IEEEtran}
%

\bibliography{main}

\vfill


\end{document}